\documentstyle[12pt]{article}
\newcommand{\rt}{\rightarrow}
\newcommand{\et}{{\it et al.}}
\begin{document}
\baselineskip 15pt
\title{\normalsize\bf PHYSICS IN CHARM ENERGY REGION } 
\author{\normalsize {Kuang-Ta Chao}\\ 
{\footnotesize\sl  CCAST (World Laboratory),~Beijing~ 100080, ~P.R.China}\\
{\footnotesize\sl Department of Physics,~Peking 
   University,~Beijing~100871,~P.R.China}}
\date{}
\maketitle    
\begin{center}
\begin{minipage}{130mm}
\begin{center}{\small ABSTRACT}\end{center}
 {Recent results on physics in the charm energy region are reviewed.
 Theoretical puzzles related to the exclusive hadronic decays of $J/\psi$
 and $\psi'$ are analyzed. New results and issues on possible glueball 
 candidates such as
 $\xi(2230), \eta(1440)$ and $f_0(1500)$ observed in $J/\psi$ radiative
 decays and other experiments are emphasized.
 Problems in charmonium production with large transverse momentum at the 
 Tevatron and fragmentation mechanisms of quarks and gluons, and in particular
 the $\psi'$ surplus and color-octet fragmentation are discussed.
 Some theoretical and experimental results in charmonium and open charm
 hadrons are also reported.}  
\end{minipage}
\end{center}
\vskip 0.3in
\section*{\normalsize\bf I. Introduction}

Physics in the charm energy region is in the boundary domain between
perturbative and nonperturbative QCD. The study of charmonium physics
has recently received renewed interest. The observed hadronic decays
of charmonium may give new challenges to the present theoretical 
understanding of the decay mechanisms.
More glueball candidates are observed in charmonium radiative decays,
and are arousing new studies of glueball physics. The observed prompt 
production of charmonium at the Tevatron and the serious disagreement
between expected and measured production cross sections
have led to new theoretical speculations about charmonium spectrum and novel 
production mechanisms. There are also many new results in
open charm physics, including new measurements of charmed meson and baryon 
decays. In this report some of the new results and the status of
theoretical studies of physics
in the charm energy region will be reviewed.

\section*{\normalsize\bf II. Problems in Charmonium Hadronic Decays}

Charmonium hadronic decays may provide useful information on understanding
the nature of quark-gluon interactions and decay mechanisms. 
They are essentially
related to both perturbative and nonperturbative QCD. The mechanism for 
exclusive hadronic decays is still poorly understood. One of the striking
observations is the so-called ``$\rho\pi$'' puzzle, i.e., in the decays 
of $J/\psi$ and $\psi'$ into $\rho\pi$ and $K^* \bar{K}$ the branching 
ratios of $\psi'$ are greatly suppressed relative to that of 
$J/\psi$ \cite{1fran}. New data from BES not only confirmed this 
observation but also found some new suppressions in the $\omega f_2$ 
and $\rho a_2$ channels \cite{li}\cite{1bes}. This gives a new challenge
to the theory of charmonium hadronic decays.

Because for any exclusive hadronic channel {\it h} the decay proceeds via the 
wave function at the origin of the $c\bar{c}$ bound state, one may expect
\begin{eqnarray}
Q_h\equiv\frac{B(\psi'\rightarrow h)}{B(J/\psi\rightarrow h)}
\approx
\frac{B(\psi'\rightarrow 3g)}{B(J/\psi\rightarrow 3g)}\approx
\frac{B(\psi'\rightarrow e^+e^-)}{B(J/\psi\rightarrow e^+e^-)}\approx 0.14.
\end{eqnarray}
Most channels like $3(\pi^+\pi^-)\pi^0$, $2(\pi^+\pi^-)\pi^0$,
$(\pi^+\pi^-) p\bar{p}$, $\pi^0 p\bar{p}$, $2(\pi^+\pi^-)$, and the newly 
measured $p\bar{p}$, $\Lambda\bar{\Lambda}$, $\Sigma^0\bar{\Sigma}^0$, 
$\Xi^-\bar{\Xi}^-$ by BES seem to approximately respect this relation. But 
$Q_h$ for $\rho\pi$ and $K^*\bar{K}$ were found (and confirmed by BES)
to be smaller by more than an order
of magnitude than the normal value 0.14. The new BES data give
\cite{li}\cite{1bes}
\begin{eqnarray}
Q_{\rho\pi}<0.0028,~~ Q_{K^{*\pm}K^\mp}<0.0048.
\end{eqnarray}
This puzzle has led to some theoretical speculations.

The Hadron Helicity Conservation theorem in QCD \cite{1bl}, suggested by 
Brodsky and Lepage, indicates that bacause vector-gluon 
coupling conserves quark helicity for massless quarks, and each hadron's
helicity is the sum of the helicity of its valence quarks, in hard process the 
total hadronic helicity is conserved (up to corrections of order $m/Q$
or higher)
\begin{eqnarray}
\sum\limits_{initial}\lambda_H = \sum\limits_{final}\lambda_H.
\end{eqnarray}
According to this theorem the decays of $J/\psi$ and $\psi'\rightarrow VP$
(vector and pseudoscalar such as $\rho\pi$ and $K^*\bar{K}$) are forbidden.
This seems to be true for $\psi'$ but not for $J/\psi$. The anomalous 
aboundance of VP states in $J/\psi$ decay then needs an explaination.

In this connection one possible solution to the $\rho\pi$ puzzle is  the 
$J/\psi-{\cal O}$ mixing 
models \cite{1nambu}\cite{1hou}\cite{1blt}. These models, 
though slightly different from each other, have the same essence that the 
enhancement of $J/\psi\rightarrow \rho\pi,~K^*\bar{K}$ is due to ${\cal O}
\rightarrow \rho\pi, ~K^*\bar{K}$, where ${\cal O}$ could be a Pomeron daughter
\cite{1nambu} or a vector glueball \cite{1hou}\cite{1blt}, which could lie in 
the region close to $J/\psi$ mass and then mixed with $J/\psi$ but not $\psi'$.

It has been suggested to search for this vector glueball in processes
$J/\psi,~\psi'\rightarrow (\eta,~\eta',~\pi\pi)+{\cal O}$, 
followed by ${\cal O}\rightarrow \rho\pi,~K^*K$. 
Obviously, the $J/\psi-{\cal O}$ mixing model depends heavily on the 
existence of a vector glueball near the $J/\psi$. It is therefore crucial to 
search for it in the vicinity of $J/\psi$. But so far there seem no signs 
for it.

Another proposed solution to this puzzle is the so-called generalized hindered 
M1 transition model \cite{1pinsky}. It is argued that because 
$J/\psi\rightarrow \eta_c\gamma$ is an allowed M1 transition while 
$\psi'\rightarrow \eta_c\gamma$ is hindered (in the nonrelativistic limit), 
using the vector-dominance 
model to relate $\psi'\rightarrow \gamma\eta_c$  
to $\psi'\rightarrow \psi\eta_c$ one could find the coupling
$G_{\psi'\psi\eta_c}$ is much smaller than 
$G_{\psi\psi\eta_c}$, and then by 
analogy, the coupling $G_{\omega'\rho\pi}$ would be much smaller 
than $G_{\omega\rho\pi}$. 
Assuming $\psi'\rightarrow \rho\pi$ to proceed via 
$\psi'-\omega'$ mixing, while $\psi\rightarrow \rho\pi$ via $\psi-\omega$
mixing, one would find that $\psi'\rightarrow \rho\pi$ is much more severely
suppressed than $\psi\rightarrow \rho\pi$.

There is another model \cite{1ct} in which a hadronic form factor is introduced
to exponentially decrease the two meson decays of $\psi'$ relative to $J/\psi$.
But this model predicts a large suppression for many two meson modes, which may
not be compatible with the present data. There is also a proposal to explain 
this puzzle based on the mechanism of sequential quark pair creation
\cite{1karl}.

Now the new BES data give a new challenge to these speculations. It is found
that in addition to $\rho\pi$ and $K^* K$ the suppression also exists in the 
VT (vector-tensor) channels of $\psi'$ decays such as $\psi'\rightarrow 
\omega f_2(1270)$ \cite{li}\cite{1bes}
\begin{eqnarray}
 Q_{\omega f_2} <0.022,
\end{eqnarray}
and the preliminary data on $\rho a_2, K^*K^*_2, \phi f'_2$ channels seem to 
also show suppressions for $\psi'$, whereas in the $b_1^\pm \pi^\mp$ channel 
there is no suppression is observed for 
$\psi'$.

The VT decays do not violate helicity conservation, therefore the suppression 
is hard to understand. Moreover, in the $J/\psi-{\cal O}$ mixing model 
the ${\cal O}\rightarrow VT$ decay is not expected to be a dominant mode, and 
therefore $J/\psi\rightarrow VT$ may not be enhanced. Moreover, using the 
vector dominance model one might relate $\psi'\rightarrow \omega f_2$ to
$\psi' \rightarrow \gamma f_2$, but the observed $\psi'\rightarrow \gamma f_2$ 
is not suppressed\cite{1lee}, and this is also confirmed by BES. 
In the generalized hindered M1 transition model, the coupling 
$G_{\omega'\omega f_2}$ for $ \omega'\rightarrow 
\omega f_2$ should not be suppressed because by analogy the coupling
$G_{\psi' \psi\chi_{c2}}$ is not small due to the fact that 
the E1 transition $\psi'\rightarrow \gamma\chi_{c2}$ is not hindered. 
Therefore via $\psi'-\omega'$ mixing
the $\psi'\rightarrow\omega'\rightarrow \omega f_2$ decay is expected to be 
not suppressed. It seems that within the scope of proposed models and 
speculations the puzzles related to the VP and VT suppressions have not 
been satisfactorily solved yet.

In order to understand the nature of these puzzles, systematical studies on 
$J/\psi$ and $\psi'$ exclusive hadronic decays are needed. Many different 
decay channels such as VP ($\rho\pi$, $K^* K$, $\omega\eta$, $\omega\eta'$, 
$\phi\eta$,     
 $\phi\eta'$,  and isospin violated $\omega\pi^0,~\rho\eta,~\rho\eta',~
\phi\pi^0,\cdots$), VT ($\omega f_2$, $\rho a_2$,  $\phi f_2$, 
$\phi f_2'$, $\cdots$)
AP($b_1\pi,\cdots$), TP ($a_2\pi, \cdots$), VS($\omega f_0,~\phi f_0,\cdots$),
VA ($\phi f_1,$  $\omega f_1$, $\cdots$) and three-body or many-body mesonic 
decays $(\omega\pi^+\pi^-,~\phi K\bar{K},\cdots)$ and baryonic decays $ 
(p\bar{p},~n\bar{n},~\Lambda\bar{\Lambda},~
\Sigma\bar{\Sigma},~\Xi\bar{\Xi},\cdots)$ are worth studying and
 may be helpful to reveal the essence of the puzzle and the nature of decay
 machnisms. In addition, to test the hadron helicity conservation theorem, 
 measurements of the decay angular momentum distribution are also
 important. E.g., it predicts a $\sin^2\theta$ distribution for
 $J/\psi,~\psi'\rightarrow \omega f_2$ \cite{1gt}.

Since the $\eta_c,~\eta'_c$ systems are the counterparts of $J/\psi,~\psi'$,
it has been suggested to study exclusive hadronic decays of $\eta_c$ and 
$\eta'_c$ \cite{1anse}\cite{1cgt}. It is argued that for any normal hadronic 
channel {\it h}, based on the same argument as for $J/\psi$ and $\psi'$, 
the following
relation should hold \cite{1cgt}
\begin{eqnarray}
P_h\equiv\frac{B(\eta'_c\rightarrow h)}
{B(\eta_c\rightarrow h)}\approx\frac{B(\eta'_c\rightarrow 2g)}
{B(\eta_c\rightarrow 2g)}\approx 1.
\end{eqnarray}
This relation differs from the ``0.14'' rule for $J/\psi,~\psi'$, because $
\eta'_c\rightarrow 2g$ is the overwhelmingly dominant decay mode, whereas for 
$\psi'$ the $\psi'\rightarrow J/\psi\pi\pi$ and $\psi'\rightarrow 
\gamma\chi_{cJ}~(J=0,1,2)$ transitions are dominant. As the ``0.14'' rule for
$J/\psi$ and $\psi'$, this relation for $\eta_c$ and $\eta'_c$ may serve
as a criterion to determine whether there exsit anomalous suppressions in the 
$\eta_c,~\eta'_c$ systems. As pointed out in \cite{1anse} 
that since the observed
$\eta_c\rightarrow VV(\rho\rho,~K^*\bar{K^*},~\phi\phi)$ and $p\bar{p}$ decays,
which are forbidden by helicity conservation, seem to be not suppressed, there
might be a $0^{-+}$ trigluonium component mixed in the $\eta_c$. It then
predicts a severe suppression for these decays of $\eta'_c$, 
which is not close to and therefore not mixed 
with the $0^{-+}$ trigluonium.  The $\eta_c$ and $\eta'_c$  hadronic decays
are being searched for at BES/BEPC, and will be studied at the $\tau$-charm
factory in the future. In this connection, it might be interesting 
to see whether
E760-E835 experiment can find $\eta'_c$ in $p\bar{p}\rightarrow\eta'_c
\rightarrow 2\gamma$. If $\eta'_c\rightarrow p\bar{p}$ is severely suppressed
by helicity conservation, as the counterpart of $\psi'\rightarrow\rho\pi$,
then it would be hopeless to  see $\eta'_c$ in $p\bar{p}$ annihilation. 
Therefore the E760-E835 experiment will further test helicity conservation 
and shed light on the extented $``\rho\pi''$ puzzle.

On the other hand, the theoretial understanding for these puzzles and, 
in general,
for the nature of exclusive hadronic decay mechanisms is still very limited.
It concerns how the $c\bar{c}$ pair convert into gluons and light quarks and, 
more importantly, how the gluons and quarks hadronize into light hadrons. 
The hadronization 
must involve long distance effects and is governed by 
nonperturbative dynamics. These problems certainly deserve a thorough 
investigation in terms of both perturbative and nonperturbative QCD.

\section*{\normalsize\bf III. Search for Glueballs in Charmonium Decays}

Existence of the non-Abelian gluon field is the key hypothesis of QCD, and 
observation of glueballs will be the most direct confirmation of the
existence of gluon field. Charmonium radiative decays into light hadrons 
proceed via $c\bar{c}\rightarrow\gamma + g + g $ and are then the gluon-rich
channels. Therefore, charmonium especially $J/\psi$ radiative decays are
very important processes in the search for glueballs. Recent experimental
studies indicate that there are at least three possible candidates of 
glueballs which are related to $J/\psi$ radiative decays.

$\bullet$~$\xi(2230)$~~~$J^{PC}=(?)^{++}$. 

The new data from BES \cite{li}
\cite{jin} confirmed the Mark III result \cite{mark} and found four decay 
modes of $\xi\rightarrow\pi^+\pi^-$, $K^+K^-$, 
$K_SK_S$, $p\bar{p}$ in $J/\psi
\rightarrow \gamma\xi$ with a narrow width of $\Gamma_\xi\approx20$ MeV.
The branching ratios are found to be
$B(J/\psi\rightarrow\gamma\xi)\times B(\xi\rightarrow X)\approx 
(5.6,3.3,2.7,1.5)\times 10^{-5}$ respectively for $X=\pi^+\pi^-, K^+K^-, 
K_SK_S, p\bar{p}$.

Combining these data with the PS $185$ experiment on $p\bar{p}\rightarrow
\xi(2230)\rightarrow K\bar{K}$ \cite{ps185}:
$B(\xi\rightarrow p\bar{p})\times B(\xi\rightarrow K\bar{K}) 
<1.5\times 10^{-4}$ (for J=2)
reveals some distinct features of the 
$\xi(2230)$: the very narrow partial decay widths to $\pi\pi$ and $K\bar{K}$
(less than $1$ MeV with banching ratios less than  $5$\%); the large production
rate in $J/\psi$ radiative decays ($B(J/\psi\rightarrow\gamma\xi)>2\times 
10^{-3}$); the flavor-symmetric couplings to $\pi\pi$ and $K\bar{K}$. These
features make $\xi(2230)$ unlikely to be a $q\bar{q}$ meson but likely to be 
a $J^{PC}=(even)^{++}$ glueball \cite{chao}\cite{hjzc}.

The $\xi(2230)$ once was interpreted as an $s\bar{s}$  meson\cite{gold}. But a 
recent quark model calculation\cite{harry} for decays of $1^{3}F_2$ and  
$1^{3}F_4$ $s\bar{s}$ mesons shows that the widths of $1^{3}F_2$ and 
$1^{3}F_4$ $s\bar{s}$ mesons are larger than $400$ MeV 
and $130$ MeV respectively.
The partial width of $1^{3}F_4$ to $K\bar{K}$ is predicted to be $(14-118)$MeV,
also much larger than that of $\xi(2230)$. Moreover, the lattice 
study of $SU(3)$ glueballs by the UKQCD group suggests the mass of $2^{++}$
glueball be $2270\pm100$MeV\cite{ukqcd}, consistent 
with the mass of $\xi(2230)$.
But the spin of $\xi(2230)$ has not been 
determined yet. ($J^{PC}=4^{++}$ will
not favor a glueball because it would require a non-S wave orbital angular
momentum between the constituent gluons, and then lead to higher mass and 
lower production rate in $J/\psi$ radiative decays than $\xi(2230)$). 
Moreover, in order to see through 
the nature of $\xi(2230)$ (e.g., by further examining 
the flavor-symmetric decays and the difference between
glueball and $q\bar{q}g$ hybrid), more data are needed for other decay modes,
such as $\eta\eta$, $\eta\eta'$, $\eta'\eta'$ and $\pi\pi\pi\pi$, 
$\pi\pi K\bar{K}$, $\rho\rho$, $K^*\bar{K^*}$, $\omega\phi$, $\phi\phi$, etc.

$\bullet$~$\eta(1440)$~~~$J^{PC}=0^{-+}$. 

For years this state has been regarded as a good candidate for the $0^{-+}$ 
glueball.
However, since both Mark III \cite{bai}
and DM2\cite{dm2} find three structures (two $0^{-+}$ and one $1^{++}$ ) in 
the energy region $1400-1500$MeV in $J/\psi\rightarrow \gamma K\bar{K}\pi$,
the status of $\iota/\eta(1440)$ as a $0^{-+}$ glueball is somewhat shaky.  
But the new (preliminary) generalized moment analysis of BES \cite{ma},
which avoids the complicated coupling effects from different intermediate
states ($K^*\bar{K}$ and $a_0\pi$), 
indicates
that the $\eta(1440)$, being one of the three structures, may have a larger
production rate in $J/\psi$ radiative decays with 
$B(J/\psi\rightarrow\gamma\eta(1440))
\cdot B(\eta(1440)\rightarrow K\bar{K}\pi)\approx2\times10^{-3}$. This may 
reinforce the $\eta(1440)$ being a $0^{-+}$ glueball candidate. 

While more data
and analyses are needed to clarify the discrepancies between Mark III, 
DM2, and BES, some theoretiacl arguments support $\iota/\eta(1440)$ being a 
$0^{-+}$ glueball. 
The helicity conservation argument favors $0^{-+}$ glueball decaying
predominantly to $K\bar{K}\pi$ \cite{chano}. Working to lowest order in 
$1/N_c$ and using chiral lagrangians also get the same 
conclusion \cite{goun}. However, the lattice QCD calculation by UKQCD predicts 
the  mass of $0^{-+}$ glueball to be $\sim 2300$MeV \cite{ukqcd}, much higher
than 1440 MeV.

$\bullet$~$f_0(1500)$~~~$J^{PC}=0^{++}$. 

The Crystal Barrel (CBAR) Collaboration
\cite{cbar} at LEAR has found $f_0(1500)$ in $p\bar{p}\rightarrow
\pi^0 f_0(1500)$ followed by $f_0(1500)\rightarrow\pi^0\pi^0,~\eta\eta,~\eta
\eta'$. This state might be the same particle as that found by WA91 
Collaboration
in central production $pp\rightarrow p_f(2\pi^+ 2\pi^-)p_s$ \cite{wa91},
and that found by GAMS in $\pi^- p\rightarrow \eta\eta{'} n,~\eta
\eta n,~4\pi^0 n$,
namely, the $G(1590)$ \cite{gams}. So far no signals have been seen in 
$J/\psi$  radiative decays in channels like $\pi\pi,~\eta\eta,~\eta\eta'$ for
$f_0(1500)$. However, it is reported recently that re-analysis of Mark III
 data on $ J/\psi\rightarrow\gamma(4\pi)$ reveals a resonance with 
 $J^{PC}=0^{++}$  at $1505$ MeV, which has a strong $\sigma\sigma$ decay mode
 \cite{bugg}. If this result is confirmed, $f_0(1500)$ may have been seen in 
 three gluon-rich processes, i.e., the $p\bar{p}$ annihilation, the central
 production with double Pomeron exchange, and the $J/\psi$ radiative decays, 
 and is therefore a good candidate for $0^{++}$ glueball. It will be 
 interesting to see whether $f_0(1500)\rightarrow\sigma\sigma\rightarrow 4\pi$
 is the main decay mode in the CBAR experiment. The mass of $f_0(1500)$ is 
 consistent with the UKQCD lattice calculation \cite{ukqcd}.

The theoretical understanding of glueballs is still rather limited. There are
different arguments regarding whether glueball decays are flavor-symmetric.

$\bullet$~Helicity Conservention.

It was argued by Chanowitz \cite{chano} that although
glueballs are $SU(3)$ flavor singlets it is inadequate to use this as a 
criterion for identifying them because large $SU(3)$ breaking may affect their
decays. In lowest order perturbation theory the decay amplitude
is expected to be proportional to the quark mass
\begin{eqnarray}
M(gg\rightarrow q\bar{q})_{J=0}\propto m_q,
\end{eqnarray}
so that decays to $s\bar{s}$ are much stronger than $u\bar{u}+d\bar{d}$ for 
$0^{++}$ and $0^{-+}$ glueballs. This is a consequence of ``helicity 
conservation''- the same reason that $\Gamma(\pi\rightarrow \mu\nu)
\gg\Gamma(\pi
\rightarrow e\nu)$, and this might explain why 
$\iota/\eta(1440)\rightarrow K\bar{K}\pi$
is dominant.

$\bullet$~Discoloring of gluons by gluons.

It was argued by Gershtein 
{\it et al.} \cite{gers} that due to QCD axial anomaly the matrix element 
$\alpha_s\!<0|G\tilde{G}|\eta'>$ gets a large value. Therefore, if the glueball
decay proceeds via production of a pair of gluons from the vacuum and 
recombination of the gluons in the initial state with the produced gluons, 
then decays into $\eta'$ will be favored. This may explain why the 
$0^{++}~ G(1590)$ has a larger decay rate into $\eta\eta'$ than $\eta\eta,~
\pi\pi,~KK$.

$\bullet$~Glueball-$q\bar{q}$ mixing.

It was argued by Amsler and 
Close \cite{ac} that for a pure glueball $G_0$ flavor democracy (equal 
gluon couplings to $u\bar{u},~d\bar{d}$ and $s\bar{s}$) will lead to the 
relative decay branching ratios $\pi\pi:K\bar{K}:\eta\eta:\eta\eta'=3:4:1:0$. 
Then by mixing with nearby $q\bar{q}$ isoscalars the mixed glueball state 
becomes
\begin{eqnarray}
|G>=|G_0> + \xi(|u\bar{u}> + |d\bar{d}> + \omega|s\bar{s}>),
\end{eqnarray}
and the observed decay branching ratios $\pi\pi:K\bar{K}:\eta\eta:\eta\eta'
=1:<0.1:0.27:0.19$ for $f_0(1500)$ may be explained in a color flux-tube 
model with certain values for mixing angles $\xi$ and $\omega$ with the nearby 
$f_0(1370)$ (an $u\bar{u}+d\bar{d}$ state) and an $s\bar{s}$ state in the 1600 
MeV region. The problem for $f_0(1500)$ and $f_0(1710)$ has
also been discussed in Ref.\cite{torn}.

$\bullet$~Resemblance to charmonium decays. 

It was argued\cite{chao} that 
pure glueball decays may bear resemblance to charmonium decays, e.g., to the 
$\chi_{c0}(0^{++})$ and $\chi_{c2}(2^{++})$ decays. Both $\chi_{c0}$ and 
$\chi_{c2}$ decays may proceed via two steps: first the $c\bar{c}$ pair 
annihilate into two gluons, and then the two gluons hadronize into light mesons
and baryons. The gluon hadronization appears to be flavor-symmetric. This is 
supported by the $\chi_{c0}$ and $\chi_{c2}$ decays, e.g., $\chi_{c0}$ is 
found to have the same decay rate to $\pi^+\pi^-$ as to $K^+K^-$, and the 
same decay rate to $\pi^+\pi^- \pi^+ \pi^- $ as to $\pi^+ \pi^-K^+K^-$,
and this is also true for $\chi_{c2}$ decays. 
For a glueball, say, a $2^{++}$ glueball, its
decay proceeds via two gluon hadronization, which is similar to the 
second step of 
the $\chi_{c2}$ decay. Therefore, a pure $2^{++}$ glueball may have 
flavor-symmetric decays. Furthermore, the $2^{++}$ glueball, if lying in the 
2230 MeV region, can only have little mixing with nearby L=3~ $2^{++}$ 
quarkonium
states, because these $q\bar{q}$ states have vanishing wave functions at the 
origin due to high angular momentum barrier which will prevent the $q\bar{q}$
pair from being annihilated 
into gluons and then mixed with the $2^{++}$ glueball. 
This might explain why 
$\xi(2230)$ has flavor-symmetric couplings to $\pi^+\pi^-$ and $K^+K^-$, if 
it is nearly a pure $2^{++}$ glueball. 
In addition, the gluon hadronization leads to many 
decay modes for $\chi_{c0}$ and $\chi_{c2}$, therefore the $2^{++}$ glueball 
may also have many decay modes. In comparison, the observed branching ratios
for $\xi(2230)\rightarrow\pi^+\pi^-,~K^+ K^-,~K_S K_S,~p\bar{p}$ may not 
exceed 6 percent. This might be very different from the conventional 
$q\bar{q}$ mesons, which usually have some dominant two-body decay modes.

Above discussions indicate that the decay pattern of glueballs could be 
rather complicated, and a deeper theoretical understanding is needed to reduce
the uncertainties. As for the glueball mass spectrum, despite the remarkable
progress made in lattice QCD calculations \cite{teper}\cite{ukqcd}\cite{ibm},
uncertainties in estimating glueball masses are still not small. For 
instance, for $0^{++}$ glueball UKQCD group gives $M=1550\pm50$MeV
\cite{ukqcd}, while IBM group gets $M=1740\pm71$MeV and $\Gamma=108\pm29$MeV
\cite{ibm}.

Another progress in the lattice calculation is the glueball matrix 
elements. For example, a calculation for $<0|Tr(g^2 G_{\mu\nu}G_{\mu\nu})|G>$
predicts a branching ratio of $5\times 10^{-3}$ in $J/\psi$ radiative decays
for $0^{++}$ glueball \cite{liang}. This may provide useful information on 
distinguishing between $f_0(1500)$ and $\theta(1720)$, or other possible 
candidates for $0^{++}$ glueball. If $f_0(1500)$ is the $ 0^{++}$ glueball,
it should have some important decay modes (e.g., $4\pi$) to 
show up in $J/\psi$
radiative decays. 

In summary, while the situation in searching for glueballs via charmonium 
decays is  very encouraging, especially with the $\tau$-charm factory in the 
future, more theoretical work should be done to make more certain predictions
on the glueball mass spectrum, the widths, the transition matrix elements, and
in particular the decay patterns.

\section*{\normalsize\bf IV. Prompt Charmonium Production at Tevatron and 
Fragmentation of Quarks and Gluons}

The study of charmonium in high energy hadron
collisions may provide an important testing ground for both perturbative
QCD and nonperturbative QCD.

In earlier calculations \cite{ruc}, 
in hadronic collisions the leading order processes 
\begin{equation}
gg\rt g\psi,~~gg,q\bar{q}\rt g\chi_c (\chi_c\rt\gamma\psi),~~
qg\rt q\chi_c (\chi_c\rt\gamma\psi),   
\end{equation}
were assumed to give dominant contributions to the cross section. But they 
could not reproduce the observed data for charmonium with large transverse 
momentum. This implies that some new production mechanisms should be important.
These are the quark fragmentation and gluon fragmentation.
\subsection*{\normalsize\bf 1. Quark Fragmentation}
In essence the quark fragmentation was first numerically evaluated in a 
calculation
for the $Z^0$ decay $Z^0\rightarrow\psi c\bar{c}$ by Barger, Cheung, and
Keung\cite{barg}(for other earlier discussions on fragmentation mechanisms
see ref.\cite{hagi}). This decay proceeds via $Z^0\rightarrow c\bar{c}$, 
followed 
by the splitting $c\rightarrow\psi c$ or $\bar{c}\rightarrow\psi\bar{c}$ (see
Fig.1), of which the rate is two orders of magnitude larger than that for 
$Z^0\rightarrow \psi gg$\cite{gub}, because the fragmentation contribution is 
enhanced by a factor of $(M_Z/{m_c})^2$ due to the fact that in fragmentation 
the charmonium ($c\bar{c}$ bound state) is produced with a seperation of order 
$1/{m_c}$ rather than $1/{m_Z}$ as in the previous short-distance processes,
e.g., $Z^0\rightarrow \psi gg$.
\vskip 4cm
\begin{center}
\begin{minipage}{120mm}
{\footnotesize  Fig.1 The quark 
fragmentation mechanism. $\psi$ is produced by 
the charm quark splitting $c \rt\!\psi c$.} 
\end{minipage}
\end{center}

These numerical calculations, which are based on the fragmentation mechanisms,
can be approximately (in the limit $m_c/{m_Z}\rightarrow 0$) re-expressed in 
a more clear and concise manner in terms of the quark fragmentation 
functions, which were
studied analytically by Chang and Chen\cite{chang}\cite{chen}, 
and by Braaten, Cheung, and Yuan\cite{bcy1}. The quark fragmentation
functions can be calculated in QCD using the Feynman diagram shown in Fig. 1. 
For instance, the fragmentation function $D_{c\rightarrow \psi}(z,\mu)$, which 
describes the probability
of a charm quark to split into the $J/\psi$ with longitudinal momentum fraction
z and at scale $\mu$, is given by\cite{bcy1}
\begin{eqnarray}
D_{c\rightarrow \psi}(z,3m_c)=\frac{8}{27\pi}\alpha_s(2m_c)^2\frac{|R(0)|^2}
 {m_c^3}\frac{z(1-z)^2(16-32z+72z^2-32z^3+5z^4)}{(2-z)^6},
\end{eqnarray} 
 where $\mu=3m_c$ and R(0) is the radial wave function at the origin of 
$J/\psi$. Large logarithms of $\mu/{m_c}$ for $\mu=O(m_Z)$ appearing in 
$D_{i\rightarrow\psi}(z,\mu)$ can be summed up by solving 
the evolution equation
\begin{eqnarray}
\mu\frac{\partial}{\partial \mu}D_{i\rightarrow\psi}(z,\mu)=\sum\limits_{j}
\int^{1}_{z}\frac{dy}{y}P_{i\rightarrow j}(z/y,\mu)D_{j\rightarrow\psi}(y,\mu),
\end{eqnarray}
where $P_{i\rightarrow j}(x,\mu)$ is the 
Altarelli-Parisi function for the splitting
of the parton of type $i$ into a parton of type $j$ with longitudinal momentum
fraction $x$. The total rate for inclusive $\psi$ production is approximately
\begin{eqnarray}
\Gamma(Z^0\rightarrow\psi+X)=2\widehat{\Gamma}(Z^0\rightarrow c\bar{c})
\int_{0}^{1}dz D_{c\rightarrow\psi}(z, 3m_c).
\end{eqnarray}
Then the branching ratio for the decay of $Z^0$ into $\psi$ relative to decay
into $c\bar{c}$ is 
\begin{eqnarray}
\frac{\Gamma(Z^0\rightarrow\psi c\bar{c})}{\Gamma(Z^0\rightarrow c\bar{c})}
=0.0234 \alpha_s(2m_c)^2\frac{|R(0)|^2}{m_c^3}\approx 2\times 10^{-4},
\end{eqnarray}
which agrees with the complete leading order calculation of $Z^0
\rightarrow\psi c\bar{c}$ in Ref.\cite{barg}.

Using the fragmentation functions, the production rates of the $B_c$ meson are
predicted  in Ref.\cite{chang}. E.g., the branching ratio of $B_c$ in $Z^0$ 
decay is about $R\approx 7.2\times 10^{-5}$ (see also Ref.\cite{chang1}).

The quark fragmentation functions to P-wave mesons have also been calculated
\cite{chen}\cite{yuan}.
\subsection*{\normalsize\bf 2. Gluon Fragmentation}
As the quark fragmentation, the gluon fragmentation may also be the dominant
production mechanism for heavy quark-antiquark bound states (e.g., charmonium) 
with large tansverse momentum. 

In previous calculations, e.g. in the gluon fusion process, charmonium states
with large $P_T$ were assumed to be produced by short distance mechanisms,
i.e., the $c$ and $\bar{c}$ are created with transverse seperations of order
$1/{P_T}$, as shown in Fig.2(a) for $gg\rightarrow\eta_c g$. However, in the 
gluon fragmentation mechamism the $\eta_c$ is produced by the gluon splitting
$g\rightarrow\eta_c g$ (while $J/\psi$ is produced by $g\rightarrow\psi gg$),
as shown in Fig.2(b).

For gluon fragmentation,
in the kinematic region where the virtual gluon and $\eta_c$ are colinear, the
propagator of this gluon is off shell only by an amount of order $m_c$, and 
enhances  the cross section by a factor of $P_T^2/{m_c^2}$. If $P_T$ is 
large enough, this will overcome the extra power of the coupling constant
$\alpha_s$, as compared with the short distance leading order process 
$gg\rightarrow\eta_c g$. The gluon fragmentation functions 
were 
calculated by Braaten and Yuan\cite{by}
\begin{eqnarray}
\int^{1}_{0} dz D_{g\rightarrow\eta_c}
(z,2m_c)=\frac{1}{72\pi}\alpha_s(2m_c)^2
\frac{|R(0)|^2}{m_c^3},
\end{eqnarray}

\begin{eqnarray}
\int^{1}_{0}dz D_{g\rightarrow\psi}(z,2m_c)=(1.2\times 10^{-3})
\alpha_s(2m_c)^3\frac{|R(0)|^2}{m_c^3}.
\end{eqnarray}
where the latter is estimated to be smaller 
than the former by almost an order of magnitude.

\vskip 5cm
\begin{center}
\begin{minipage}{120mm}
{\footnotesize 
Fig.2(a) A Feynman diagram for $gg\rt c\bar{c}g$ that contributes to $\eta_c$ 
 production at order $\alpha_s^3$;~~ 
 Fig.2(b) A Feynman  diagram for 
 $gg\rt c\bar{c}gg$ that contributes to $\eta_c$ production at order 
 $\alpha_s^4$. For the virtual gluon at large $P_T$, with~$q_0=O(P_T),~
 q^2=O(m_c^2)$, the contribution is dominant.} 
\end{minipage}
\end{center}

The gluon fragmentation into P-wave heavy quarkonium was also 
studied\cite{by1}. The P-wave state (e.g., $\chi_{cJ}$)   can arise from two 
sources i.e. the production of a color-singlet 
P-wave state, and the production 
of a $c\bar{c}$ pair in a color-octet S-wave state, which is then projected
onto the $\chi_{cJ}$ wave functions. With two parameters which characterize
the long-distance effects, i.e., the derivative of P-wave wavefunction at the
origin and the probability for an S-wave color-octet $c\bar{c}$ pair in
the color-singlet $\chi_{cJ}$ bound state, the fragmentation probabilities
for a high transverse momentum gluon to split into $\chi_{c0},~\chi_{c1},~ 
\chi_{c2}$ are estimated to be $0.4\times10^{-4},~1.8\times 10^{-4},~
2.4\times10^{-4},$ respectively\cite{by1}. They could be the main source 
of $\chi_{cJ}$ production at large $P_T$ in $p\bar{p}$ colliders.

Since fragmentating gluons are approximately transverse, their products are
significantly polarized. Cho, Wise, and Trivedi\cite{cwt} find that in gluon
fragmentation to $\chi_{cJ}(1P)$ followed by $\chi_{cJ}\rightarrow
\gamma J/\psi$ the helicity levels of $\chi_{c1},~\chi_{c2}$, and $J/\psi $
are populated according to certain ratios, e.g., $D_{\chi_{c1}}^{h=0}:
D_{\chi_{c1}}^{|h|=1}\approx 1:1$, $D_{J/\psi}^{h=0}:D_{J/\psi}^{|h|=1}\approx
1:3.4$.

The gluon fragmentation to $J^{PC}=2^{-+}~~^1D_2$ 
quarkonia was also studied,
and these D-wave state's polarized fragmentation functions 
were computed \cite{wise}.

\subsection*{\normalsize\bf 3. The $\psi'$ surplus problem at the Tevatron}

In 1994, theoretical calculations were compared with data on inclusive 
$J/\psi$ and $\psi'$ production at large transverse momentum at the Tevatron
\cite{cdf}, where large production cross sections were observed. The 
calculations include both the conventional leading order mechanisms and the 
charm and gluon fragmentation contributions \cite{cg}\cite{bdfm}.

For $\psi$ production both fragmentation 
directly into $\psi$ and fragmentation 
into $\chi_c$ followed by the radiative decay $\chi_c \rt \psi + \gamma$ are 
considered. Fragmentation functions for 
$g \rt \psi, ~c \rt \psi,~g \rt \chi_c,~c \rt \chi_c,$ and $\gamma\rt\psi$ 
are used.

These calculations indicate that

(1) Fragmentation dominates over the leading-order mechanisms for $P_T >5$ GeV.

(2) The dominant production mechanism by an order of magnitude is gluon
fragmentation into $\chi_c$ followed by $\chi_c\rt\gamma\psi$.

For $\psi'$ production the fragmentaions $g \rt \psi',~c \rt \psi',
 ~\gamma \rt\psi'$ and the leading order mechanisms are included but no 
contribution from any higher charmonium states is taken 
into consideration. The 
dominant production mechanisms are gluon-gluon fusion for $P_T<5$GeV, and
charm quark fragmentaion into $\psi'$ for large $P_T$. However, the calculated
production cross section of $\psi'$ is too small by more than an order of 
magnitude (roughly a factor of 30) \cite{bdfm}\cite{rs1}. 
This serious disagreement, the so-called $\psi'$ surplus problem, has caused 
many theoretical speculations.

The radically excited $2^3P_{1,2}~(\chi_{c1}(2P)$ and $\chi_{c2}(2P))$ states
have been suggested to explain the $\psi'$ surplus problem \cite{cwt}
\cite{rs2}\cite{close}. These states can be produced via gluon and charm 
fragmentaion as well as the conventional gluon fusion mechanism, and then 
decay  into $\gamma\psi'$ through E1 transitions. Large branching ratios of
$B(\chi_{cJ}(2P)\rt\psi'(2S)+\gamma)=(5\sim10)\%$ (J=1, 2) are required to 
explain the $\psi'$ production enhancement. Within the potential model with
linear confinment, the masses of these 2P states are predicted to be, e.g., 
$ M(\chi_{c0}(2P))=3920 MeV,~M(\chi_{c1}(2P))=3950 MeV,$ and 
$M(\chi_{c2}(2P))=3980 MeV$ \cite{isgur}, therefore OZI-allowed hadronic 
decays  like $\chi_{c0}(2P)\rt D\bar{D}$,~$ \chi_{c1}(2P)\rt D^*\bar{D}+c.c.,$
and $\chi_{c2}(2P)\rt D\bar{D},~D^*\bar{D}+c.c.$, can occur. It is not clear
whether these hadronic widths are narrow, making the brancing ratios 
$B(\chi_{cJ}(2P)\rt \psi'(2S)+\gamma)$ large enough to explain the $\psi'$ 
production data.

One possibility is that since decays 
$\chi_{c2}(2P)\rt D\bar{D},~D^*\bar{D}+c.c.$
proceed via $L=2$ partial waves, they could be suppressed \cite{cwt}.
These OZI-allowed hadronic decays are estimated in a flux-tube model and they
 could be further suppressed (aside from the D-wave phase space for 
 $\chi_{c2}(2P))$ due to the node structure in the radial wave functions of
 excited states \cite{page}. With suitble parameters used, the widths of 
 $\chi_{c2}(2P)$ and $\chi_{c1}(2P)$ could be as narrow as 
 $\Gamma\approx(1\sim 10)$MeV.

 There is another possibility that the $\chi_{c1}(2P)$ could lie bellow the 
 $D^*\bar D$  threshold, and then with roughly estimated $\Gamma(\chi_
 {c1}(2P)\rt $ light hadrons)$\approx 640$ KeV, $\Gamma(\chi_{c1}(2P)\rt 
 \gamma \psi') \approx 85$ KeV, one could get $B(\chi_{c1}(2P)\rt 
 \gamma \psi')\approx 12\%$ \cite{cq}. This 
 possibility relies on the expectation 
 that color screening effect of light quark pair on the heavy $Q$-$\bar{Q}$ 
 potential, observed in lattice QCD calculations, would lead to a screened 
 confinement potential which makes the level spacings of excited $c\bar{c}$ 
 states lower than that obtained using the linear potential (e.g., 
 $\psi(4160)$ and $\psi(4415)$ could be 4S and 5S rather than 2D and 4S 
 states, respectively).

 Moreover, it is suggested that the $c\bar{c}g$ hybrid states could make a 
 significant contribution to $J/\psi$ and $\psi'$ signals at the
 Tevatron\cite{close}, since the color octet production mechanism is 
 expected to be important, and hybrid states contain gluonic excitations in
 which the $c\bar{c}$ are in the S-wave color octet configuration. In 
 particular, the negative parity hybrid states, including $(0, 1, 2)^{-+},~ 
 1^{-+}$, lying in the range 
 $4.2\pm 0.2$GeV, could be a copious source of $J/\psi$ and $\psi'$, 
 through radiative and hadronic transitions.
\subsection*{\normalsize\bf 4. Color Octet Fragmentation Mechanism}

Baseed on a general factorization analysis of the annihilation and
production of heavy quarkonium \cite{bbl}\cite{lepage}, 
Braaten and Fleming proposed a
new mechanism, i.e. the color-octet fragmentation 
for the $J/\psi$ and $\psi'$ production at large $P_T$
\cite{bf}.

In the framework of NRQCD theory \cite{lepage}\cite{bbl}, 
which is based on a double power series expansion 
in the strong coupling constant $\alpha_s$ and the small velocity parameter
$v$ of heavy quark, the fragmentaion functions can be factored into 
short-distance coefficients and long-distance factors that contain
all the nonperturbative dynamics of the formation of a bound state containing
the $c\bar{c}$ pair. E.g., for $g\!\rt\!\psi$ fragmentation 
\begin{equation}
D_{g\rt\psi}(z,\mu)=\sum\limits_{n}d_n(z,\mu)<0|{\cal O}_n^{\psi}|0>,
\end{equation}
where ${\cal O}_n^{\psi}$ are local four fermion (quark) operators.

For the physical $\psi$ state, the wavefunction can be expressed as Fork 
state  decompositions which include dynamical gluons and color-octet
$(Q\bar Q)_8$ components
\begin{eqnarray}
|\psi> &=& O(1)|(Q\bar{Q})_1(^{3}S_1)>+O(v)|(Q\bar{Q})_8(^{3}P_J)g>  
\nonumber \\ 
 &+& O(v^2)|(Q\bar{Q})_8(^{3}S_1)gg> + \cdots .
\end{eqnarray}
Therefore there are two mechanisms for gluon fragmentation into $\psi$:

(1)Color-singlet fragmentation $g^*\rt\!c\bar{c}gg.$ 
Here $c\bar{c}$ is produced in a color-singlet $^3S_1$ state.
The matrix element $<{\cal O}^{\psi}_1(^3S_1)>$ is of order $m_c^3v^3$, which
is related to the Fork state $|(c\bar{c})_1(^3S_1)>$ in $\psi$, so the 
contribution to fragmentation function is of order $\alpha_s^3v^3$.

(2)Color-octet fragmentation $g^*\rt c\bar{c}.$ 
Here $c\bar{c}$ is produced in a color-octet$~^3S_1$ state.
The matrix element $<{\cal O}^{\psi}_8(^3S_1)>$  is of order $m_c^3v^7$, 
which is related to the Fork state 
$|(c\bar{c})_8(^3S_1)gg>$ in $\psi$, so the contribution
to fragmentation function is of order $\alpha_s v^7$.

It is clear that the color-octet fragmentation $g^*\rt c\bar{c}$ is enhanced
by a factor of $\sim\alpha_s^{-2}$ from the short-distance coefficients, and 
suppressed by a factor of $\sim v^4$ from the long-distance matrix elements,
as compared with the color-singlet fragmentation. Since for charmonium 
$v^2\approx 0.25\sim 0.30$ is not very small, the color-octet fragmentation 
could be dominant in some cases, e.g., in the $\psi'$ production at
large transverse momentum.

In the case of $\psi'$, if the observed large cross section is really
due to color-octet 
fragmentation the matrix element $<{\cal O}^{\psi'}_8(^3S_1)>$ will be 
determined by fitting the CDF data on the $\psi'$ production rate at  large
$P_T$ 
\begin{eqnarray}
<{\cal O}^{\psi'}_8(^3S_1)> =0.0042 GeV^3,
\end{eqnarray}
while the color-singlet matrix element $<{\cal O}^{\psi'}_1(^3S_1)>$
is determined by the $\psi'$ leptonic decay width which is related to 
the wave function at the origin
\begin{eqnarray}
<{\cal O}^{\psi'}_1(^3S_1)>\approx \frac{3}{2\pi}|R_{\psi'}|^2=0.11 GeV^3.
\end{eqnarray}
The color-octet matrix element is smaller by a factor of 25 than the
color-singlet matrix element, consistent with 
suppression  by $v^4$. Therefore the color-octet fragmentation could be a
possible solution to the $\psi'$ surplus problem.

The color-octet fragmentation will also make a substantial contribution to
the $J/\psi$ production at large $P_T$, and may compete with gluon 
fragmentation 
into $\chi_{cJ}$ followed by $\chi_{cJ}\rt\gamma J/\psi$.

The color-octet fragmentation mechanism might be supported by 
the new data from CDF\cite{cdf} and D0\cite{d0} at the Tevatron. 
New results for the fraction of $J/\psi$ which come from 
the radiative decay of 
$\chi_c$ are (see Fig.3 for the CDF result)
\begin{eqnarray}
CDF:~~f_{\chi}^{J/\psi}=
(32.3\pm 2.0\pm 8.5)\%~~(P_T^{J/\psi}>4GeV,~|\eta^{J/\psi}|<0.6),
\end{eqnarray}
\begin{eqnarray}
D0:~~~f_{\chi}^{J/\psi}=
(30\pm 7\pm 10)\%~~(P_T^{J/\psi}>8GeV,~|\eta^{J/\psi}|<0.6).
\end{eqnarray}

\vskip 5.5cm
\begin{center}
\begin{minipage}{120mm}
{\footnotesize  Fig.3 The fraction of $J/\psi$ from $\chi_c$ as a function
of $P_T^{J/\psi}$ with the contribution from $b$ quark's removed, 
measured by CDF  \cite{cdfc}.}
\end{minipage}
\end{center}
This implies that the majority of prompt $J/\psi$ at large $P_T$ do not 
come from $\chi_c$, and the gluon fragmentation into $\chi_c$ is not the 
dominant mechanism for $J/\psi$ production at large $P_T$. 
The observed production cross section of $J/\psi$ from $\chi_c$ is  in
reasonable agreement with the theoretical calculations while the direct
$J/\psi$ production cross section is a large factor above the prediction
(see Fig.4 for the CDF result).

\vskip 11cm
\begin{center}
\begin{minipage}{120mm}
{\footnotesize  Fig.4 Differential cross sections of prompt $J/\psi$ as
a function
of $P_T^{J/\psi}$ with the contribution from $b$ quark's  removed,
measured by CDF \cite{cdfc}.
The dotted curve repesents the total fragmentation contribution (but
without the color-octet fragmentation), and the dashed curve represents
the leading-order contribution \cite{bdfm}.} 
\end{minipage}
\end{center}
Although
this result might favor the color-octet fragmentation mechanism for the
direct production of $J/\psi$,
it is still premature to 
claim that it is the real source of $J/\psi$ production.

In order to further test the color-octet fragmentation mechanism for the 
production of $J/\psi$ and, in particular, $\psi'$ at the Tevatron, some
studies are required. First, the produced $\psi'$ should be transversely 
polarized \cite{wise}, and the experimental observation of a large 
transverse $\psi'$ spin alignment would provide strong support for the 
color-octet production mechanism of $\psi'$. Another important test is
to  apply the same mechanism to the $b\bar {b}$ systems.
The integrated and differential production cross sections for the 
$\Upsilon(1S), \Upsilon(2S), \Upsilon(3S)$ have been measured by both
CDF \cite{cdfb} and D0 \cite{d0}. The production rates are generally found
to be higher than that with color-singlet fragmentations. The color-octet
production mechanism does help to explain some of the discrepancies
\cite{cl}.  

In this connection it is worthwhile to note that the problem of $J/\psi$
and especially $\psi'$ surplus production has also been observed by the
fixed target experiment (e.g., in 800~GeV proton-gold collisions) \cite{fix}.
In collisions at lower energies, fragmentation is not expected to be
dominant. It is not clear whether the $\psi'$ surplus observed both at
the Tevatron and fixed-target (with the same enhancement factor of
about 25 relative to the expected production rates) has the same origin
or not. Further experimental and theoretical investigations are needed.

\section*{\normalsize\bf V. Some Results in Charmonium and Open Charm Physics}

Here some theoretical results on charmonium and open-charm hadrons
are reported.

{\bf $\bullet$}~~The $Q\bar{Q}$ spin dependent potential. 

There have been many discussions about the $Q\bar{Q}$ spin dependent 
potential (see e.g. \cite{ef}\cite{g}\cite{bnt}).  
A new formula for the heavy-quark-antiquark spin-dependent potential is 
given using the techniques developed in heavy-quark effective theory 
\cite{cko}. The leading logarithmic quark mass terms
emerging from the loop contributions are explicitly extracted and summed up.
There is no renormalization scale ambiguity in this new formula. The 
spin-dependent potential in the new formula is expressed in terms of three
independent color-electric and color-magnetic field correlation functions,
and it includes both the Eichten-Feinberg formula \cite{ef}\cite{g}
and one-loop QCD result \cite{bnt} as special cases. 
For hyperfine splittings with $\Lambda_{\overline{MS}}=200-500 MeV$, 
the new formula gives \cite{co}
$M(J/\psi)-M(\eta_c)\approx 110-120 MeV $,
$M(\Upsilon)-M(\eta_b)\approx 45-50 MeV $, and
\begin{equation} 
M(^1P_1)-M(^3P_J)\approx 2-4 MeV 
\end{equation}
for $c\bar{c}$, which is larger
than the present E760 result $(\sim 0.9 MeV)$ \cite{e760}, and other 
theoretical
predictions (e.g. \cite{halzen}). But this tiny mass difference may be 
sensitive to other effects, e.g., the coupled-channel mass shifts.

A set of general relations between the spin-independent and spin-dependent
potentials of heavy quark and antiquark interactions are derived from 
reparameterization invariance in the Heavy Quark Effective Theory \cite{ck}. 
They are useful in understanding the spin-independent and
spin-dependent relativistic corrections to the leading order
nonrelativistic potential.

{\bf $\bullet$}~~Relativistic corrections to $Q\bar{Q}$ decay widths and
the determination of $\alpha_s(m_Q)$.

Charmonium mass spectrum and decay rates can be very useful in determining
the QCD coupling constant $\alpha_s$. In recent years remarkable progresses
have been made in lattice calculations \cite{latticec}\cite{shig}. 
On the other hand,
many decay processes may be subject to substantial relativistic corrections,
making the determination of $\alpha_s$ quite uncertain 
\cite{ml}\cite{kwong}.

The decay rates of $V\rightarrow 3g$ and $V\rightarrow e^+ e^-$
for $V=J/\psi$ and $\Upsilon$ may be expressed in terms of the Bethe-Salpeter
amplitudes, and to the first order relativistic correction and QCD radiative
correction it is found that \cite{chl}
\begin{eqnarray}
\Gamma(V\rightarrow e^+e^-)=\frac{4\pi {\alpha}^2 e_Q^2}{m_Q^2}
|\int d^3 q (1-\frac{2{\vec q}^2}{3m_Q^2})\psi_{Sch}(\vec q)|^2%
(1-\frac{16\alpha_s}{3\pi}), \nonumber \\
\Gamma(V\rightarrow 3g)=\frac{40({\pi}^2-9)\alpha_s^3(m_Q)}{81m_Q^2}
|\int d^3 q[1-2.95{{\vec q}^2\over{m_Q^2}}]
\psi_{Sch}(\vec q)|^2 (1-\frac{S_Q\alpha_s}{\pi}),
\end{eqnarray}
where $S_c=3.7,~ S_b=4.9$ (defined in the $\overline{MS}$ scheme at the
heavy quark mass scale) \cite{ml}\cite{kwong}.
This result shows explicitly that the relativistic correction 
suppresses the gluonic 
decay much more severely than the leptonic decay.
Using the meson wavefunctions obtained by solving the BS equation
with a QCD-inspired interquark potential, and the experimental values 
of decay rates \cite{pdg}, it is found that\cite{chl}
\begin{equation} 
\alpha_{s}(m_c)=0.26-0.29,~~~  \alpha_s(m_b)=0.19-0.21,
\end{equation}
at $m_c=1.5~GeV$ and $m_b=4.9~GeV$.
These values for the QCD coupling
constant are substantially enhanced, as compared with the ones obtained 
without relativistic corrections. However, it should be emphasized that
these numerical results can only serve
as an improved estimate rather than a precise determination, due to large 
theoretical uncertainties related to the scheme dependence of QCD radiative 
corrections \cite{blm} and higher order relativistic corrections. 
This result is consistent with that obtained using finite size 
vertex corrections 
\cite{chiang}.

{\bf $\bullet$}~~Heavy meson decay constants. 
 
Discussions on the heavy meson decay constants are very extensive. 
In the framework of heavy quark effective theory (HQET), QCD sum rules
are used to estimate the nonperturbative effects 
\cite{neu}\cite{ele}\cite{ball}\cite{hl}.
The first systematic investigation was given in \cite{neu}, and a further 
improvement was obtained by seperating the subleading order from the
leading one \cite{ball}.

In a recent work \cite{hl1} 
the SU(3) breaking effects in the leading and subleading parameters 
appeared in the heavy quark expansion of decay constans of the heavy-light
mesons are systematically analyzed to two loops accuracy using QCD sum rules.
It is found that the SU(3) breaking effects in the decay constant 
of the pseudoscalar are respectively 
\begin{equation} 
f_{B_s}/f_B=1.17\pm 0.03,~~~~~f_{D_s}/f_D=1.13\pm 0.03.
\end{equation}
These results are in agreement with recent lattice QCD calculations
\cite{shig}. In addition, the 
ratios of vector to pseudoscalar meson decay constants are found to be
\begin{equation} 
f_{B_s^*}/f_{B_s}=f_{B_s}/f_B=1.05\pm 0.02,
\end{equation}
 and the SU(3) breaking effect 
in the mass is about $82\pm 8$MeV. 

Another approach to estimating nonperturbative effects on heavy mesons
is to combine HQET with chiral perturbation theory \cite{yan}.
In the framework of the heavy-light chiral perturbation theory
(HLCPT) the heavy meson decay constants are discussed \cite{gjms}
and the effects of excited states on the chiral loop corrections
are further considered \cite{falk}. 

In a recent work the vector meson contributions are introduced in 
HLCPT and the lagrangian and current to the order 
$1/{m_Q}$ are constructed \cite{dghj}. 
With this, to the order $1/{\Lambda_{csb}^2}$ ($\Lambda_{csb}$ is 
the chiral symmetry breaking scale), corrections to $f_D$ and 
$f_B$ arising from coupled-channel effects to order $1/{m_c} $ and $1/{m_b}$
are calculated.
At the tree level in HLCPT, using the relativistic B-S equation with kernel 
containing a confinement term and a gluon exchange term in a covariant 
generalization of the Coulomb gauge \cite{dai}, the decay constants 
$f_D^{(0)}$ and $f_B^{(0)}$
when $m_Q\rt \infty $ as well as the $1/{m_Q}$ corrections are calculated.  
HLCPT and the heavy quark effective theory (HQET) are matched at the 
scale $\Lambda_{csb}$. Adding the perturbative and 
nonperturbative contributions
the values for $f_{D}$ and $f_{B}$ are found to be
\begin{equation} 
f_D\approx f_B\approx 200~MeV,
\end{equation}
which is in agreement with lattice calculations
\cite{shig}.

We now turn to some new experimental results in open charm physics.
The CLEO Collaboration has given following results.

{\bf $\bullet$}~~More accurate or the first measurements of $D^0$ decays
\cite{c1}.

\begin{center}
\begin{tabular}{|c|c|c|}\hline
Channel & B($\%$) & PDG($\%$)\\\hline
$K^+K^-$ & 0.455$\pm$0.029$\pm$0.032 & 0.454$\pm$0.029\\\hline
$K^0\bar{K}^0$ & 0.048$\pm$0.012$\pm$0.013 & 0.11$\pm$0.04\\\hline
$K^0_SK^0_SK^0_S$ & 0.074$\pm$0.010$\pm$0.018 & 0.089$\pm$0.025\\\hline
$K^0_SK^0_S\pi^0$ & $<$0.063 $at$ 90 CL$\%$ &\\\hline
$K^+K^-\pi^0$ & 0.107$\pm$0.030 &\\\hline
\end{tabular}
\end{center}

The theoretical prediction for $B(K^+K^-)$ is in the range 0.14-0.6, and for
$B(K^0\bar{K}^0)$ is 0-0.3.

{\bf $\bullet$}~~Observation of the Cabibbo suppressed charmed baryon decay of
$\Lambda^+_c\rightarrow p\phi$ and $pK^+K^-$, 
compared with $\Lambda^+_c\rightarrow pK^-\pi^+$ \cite{c2}. 
\begin{equation}
B(p\phi)/B(pK\pi)=0.024\pm 0.006\pm 0.003,
\end{equation}
\begin{equation}
B(pKK)/B(pK\pi)=0.039\pm 0.009\pm 0.007,
\end{equation}
\begin{equation}
B(p\phi)/B(pKK)=0.62\pm 0.20\pm 0.12.
\end{equation}

The theoretical predictions range from 0.01 to 0.05 for $B(p\phi)/B(pK\pi)$.

{\bf $\bullet$}~~Measurement of the isospin-violating decay 
$D^{*+}_s\rightarrow D^+_s\pi^0$ \cite{c3}.
\begin{equation}
\frac{\Gamma(D^{*+}_s\rightarrow D^+_s\pi^0)}{\Gamma(D^{*+}_s\rightarrow
D^+_s\gamma)}=0.062^{+0.020}_{-0.018}\pm 0.022.
\end{equation}

This isospin-violating decay is expected to proceed through OZI-allowed
decay $D^{*+}_s\rightarrow D^+_s\eta$ (via the $s\bar{s}$ component in $\eta$) 
and the $\eta-\pi^0$ mixing \cite{chod}. This decay also implies that 
$D^{*+}_s$ has natural spin-parity (most likely $1^-$).

{\bf $\bullet$}~~Measurement of the relative branching ratios of $D^+_s$ 
to $\eta e^+\nu$ and $\eta^{\prime}e^+\nu$, compared to $\phi e^+\nu$
\cite{c4}.
\begin{equation}
\frac{B(D^+_s\rightarrow\eta e^+\nu)}{B(D^+_s\rightarrow\phi e^+\nu)}
=1.24\pm 0.12\pm 0.15.
\end{equation}
\begin{equation}
\frac{B(D^+_s\rightarrow\eta' e^+\nu)}{B(D^+_s\rightarrow\phi e^+\nu)}
=0.43\pm 0.11\pm 0.07.
\end{equation}
These results favor the prediction of the ISGW2 model \cite{isgw2}.

{\bf $\bullet$}~~Measurement of $\Xi^+_c$ decay branching ratios relative to
$\Xi^+_c\rightarrow\Xi^-\pi^+\pi^+$ \cite{c5}.
\begin{center}
\begin{tabular}{|c|c|c|}\hline
Decay Mode & events & $B/B(\Xi^+_c\rt\Xi^-\pi^+\pi^+)$\\\hline
$\Sigma^+K^-\pi^+$ & 119$\pm$23 & 1.18$\pm$0.26$\pm$0.17\\\hline
$\Sigma^+K^{*0}$ & 61$\pm$17 & 0.92$\pm$0.27$\pm$0.14\\\hline
$\Lambda K^-\pi^+\pi^+$ & 61$\pm$15 & 0.58$\pm$0.16$\pm$0.07\\\hline
$\Theta^-\pi^+\pi^+$ & 131$\pm$14 & 1.0\\\hline
\end{tabular}
\end{center}

There are also some experimental results from the ARGUS Collaboration.

{\bf $\bullet$}~~Leptonic branching ratios of $D^0$ \cite{a1}. 
\begin{equation}
B(D^0\rightarrow e^+\nu_eX)=6.9\pm 0.3\pm 0.5 \%,
\end{equation}
\begin{equation}
B(D^0\rightarrow \mu^+\nu_{\mu}X)=6.0\pm 0.7\pm 1.2 \%.
\end{equation}
These values are smaller than the world average values \cite{pdg}.

{\bf $\bullet$}~~Measurement of the decay $D_{s2}^+(2573)\rt D^0K^+$
\cite{a2}. The observed mass and width $\Gamma=(10.4\pm 8.3\pm 3.0)~
MeV$ of this resonance are consistent with that obtained by CLEO.

{\bf $\bullet$}~~Evidence for the $\Lambda_c^{*+}(2593)$ production \cite{a3}.

Finally, BES Collaboration has reported the leptonic branching ratio of
$D_s$ using $(148\pm 18\pm 13)~D_s$ events \cite{bs}.
\begin{equation}
B(D^+_s\rightarrow e^+\nu_eX)=(10.0^{+6.5+1.3}_{-4.6-1.2}) \%.
\end{equation}

\section*{\normalsize\bf VI. Conclusions}
While impressive progress in experiment
has been made in physics in the charm energy region, some
theoretical issues need to be clarified. 

The new data and puzzles in exclusive hadronic decays of $J/\psi$
and $\psi'$ give new challenges to the theory of hadronic decays.

With the new observation for $\xi(2230)$ and $f_0(1500)$, the situation
in searching for glueballs is encouraging, but theoretical
uncertainties related to the properties of glueballs still remain and need
to be further reduced. 

For the prompt production of charmonium at large transverse
momentum, gluon and quark fragmentations dominate over leading-order
parton fusions. Color-singlet fragmentation is not the dominant mechanism
for $J/\psi$ and $\psi'$ production. Color-octet fragmentation seems to
be important to explain the $J/\psi$ and, in particular, the $\psi'$
excess, but further tests are required. The mechanism of charmonium 
production at fixed-target also needs studying.  

The study of open charm physics is in continuous progress.
This is important for testing the Standard Model and understanding
both perturbative and nonperturbative QCD.

In the future, with new experiments at $e^+e^-$ colliders, hadronic 
colliders, fixed
target, and, in particular, at the proposed $\tau$-charm factory, and
with the theoretical progress in lattice QCD and other nonperturbative
methods, a deeper understanding of physics in the charm energy region 
will be achieved.

\section*{\normalsize\bf Acknowledgements}
I would like to thank my colleagues, in particular, Y.B. Dai, 
Z.X. He, T. Huang,
Y.P. Kuang, J.M. Wu, and H. Yu for many very helpful discussions
and suggestions. I would also like to thank Y.F. Gu, S. Jin, J. Li, W.G. Li, 
 C.C. Zhang, Z.P. Zheng, and Y.C. Zhu for useful discussions 
on experimental results. Thanks are also due to H.W. Huang, J.F. Liu, Y. Luo, 
and especially C.F. Qiao for their help in preparing this report.
I also wish to thank V. Paradimitriou for providing me with the new
data from CDF.

\end{document}